\documentclass[12pt]{article}
\usepackage{a4wide}
\usepackage{amssymb}
\usepackage{amsmath}
\usepackage{graphicx}
\usepackage{mathdots}

\def\makeatletter{\catcode`\@=11}
\makeatletter
\def\mathbox#1{\hbox{$\m@th#1$}}%
\def\math@ccstyles#1#2#3#4#5#6#7{{\leavevmode
      \setbox0\mathbox{#6#7}%
      \setbox2\mathbox{#4#5}%
      \dimen@ #3%
      \baselineskip\z@\lineskiplimit#1\lineskip\z@
      \vbox{\ialign{##\crcr
             \hfil \kern #2\box2 \hfil\crcr
             \noalign{\kern\dimen@}%
             \hfil\box0\hfil\crcr}}}}
\def\mathaccstyles{\math@ccstyles\maxdimen}
\def\maththroughstyles{\math@ccstyles{-\maxdimen}}
\def\unity%
 {\maththroughstyles{.45\ht0}\z@\displaystyle {\mathchar"006C}\displaystyle 1}

\begin{document}

\centerline{\LARGE \bf Gauge/gravity duality and RG flows }
\vspace{1truecm}
\centerline{\LARGE \bf in 5d gauge theories}



\vspace{2truecm}

\centerline{
    {\large \bf Alessandro Pini ${}^{a}$} \footnote{pinialessandro@uniovi.es}
     {\bf and}
    {\large \bf Diego Rodr\'{\i}guez-G\'omez${}^{a}$} \footnote{d.rodriguez.gomez@uniovi.es}}

\vspace{1cm}
\centerline{{\it ${}^a$ Department of Physics, Universidad de Oviedo}} \centerline{{\it Avda.~Calvo Sotelo 18, 33007, Oviedo, Spain}}
\vspace{2cm}

\centerline{\bf ABSTRACT}
\vspace{1truecm}

\noindent We discuss RG flows in 5d gauge theories triggered either by VEV's of mesonic or baryonic operators. As a warm-up, we explicitly discuss the counterpart of these flows in 4d gauge theories with $\mathcal{N}=2$ supersymmetry by focusing on the $A_1$ theory. As opposed to the $\mathcal{N}=1$ case, in cases with 8 supercharges we need to solve a more involved PDE. In the 5d case the boundary conditions for such equation play a crucial role in order to reproduce the expected spectrum.

\newpage

\tableofcontents

\section{Introduction}

The basic version of the $AdS_{d+1}/CFT_d$ correspondence equates gravity in an $AdS_{d+1}$ background with a certain $CFT_d$ living on its boundary. On general grounds, the $AdS_{d+1}$ space arises as the near brane geometry of a stack of branes at the tip of a certain --typically singular-- cone whose radial coordinate becomes the $AdS_{d+1}$ radius while its base encodes the details of the $CFT_d$. While the $CFT_d$ might have a rich structure of vacua, the $AdS_{d+1}$ dual describes just the trivial one where no operator is taking a VEV. In the following we will assume the $CFT_d$ to have a moduli space of vacua. Hence, it is natural to probe its structure by moving among vacua upon considering operator VEV's in the $CFT_d$. On general grounds, operator VEV's will trigger RG flows from the original theory to a new IR fixed point. Conversely the dual gravitational description will not anymore be exactly $AdS_{d+1}$ but only an asymptotically $AdS_{d+1}$. In fact, since we are considering moving on the moduli space of the $CFT_d$ rather than adding a deformation, the asymptotics should be just the same. Then, the internal structure of the space as one goes from the boundary to the bulk encodes the precise details of the RG flow, in particular which operators are taking a VEV as well as their dimensions and other quantum numbers. Of course, deep in the interior, one expects a different $AdS_{d+1}$ throat to locally develop standing for the IR fixed point theory.

In this paper we study aspects of these RG flows in theories with 8 supercharges in 4 and 5 dimensions. Our flows come in two broad types, namely one in which a meson-like operator acquires a VEV, and another in which a baryon-like operator acquires a VEV. The former corresponds to locating the branes sourcing the geometry away from the tip of the singular cone --hence we dub them singular flows-- while the later correspond to blow-ups of the cone --we refer to them as baryonic flows--. As the name indicates, in the baryonic flows a baryonic symmetry undergoes spontaneous breaking by the VEV.  In the 4d case, similar flows have been considered, mostly for the $\mathcal{N}=1$ case (see \textit{e.g.} \cite{Klebanov:2007us,Chen:2007em,Martelli:2007mk,Klebanov:2007cx,Krishnan:2008kv,Martelli:2008cm}). Here we consider in detail the $\mathcal{N}=2$ case, which presents some particularities such as the need to solve a rather involved PDE. Furthermore, this case serves as warm-up for the basically unexplored 5d case. In the later we find an interesting interplay among boundary conditions which allows to find the correct dimensions for the operators in the gravity side.

We will particularize our discussion to branes probing the $A_1$ singularity. While the 4d case corresponds to D3 branes probing  $\mathbb{C}^2/\mathbb{Z}_2\times \mathbb{C}$, the 5d case corresponds to D4 branes probing the $A_1$ singularity wrapped by an $O8^-$ plane with $N_f$ D8. Then, the organization of the paper is as follows. In section \ref{4d} we study the gravitational aspects of $d=4$ flows, both in the singular and resolved cases. We then turn to the gauge theory description, discussing first the wavefunctions on the $A_1$ singularity. Through these we can explicitly identify the dual operators, both in the singular and resolved case. The latter nicely fits as a broken baryon symmetry phase. Indeed, using standard techniques, we identify both the VEV of the baryon condensate and the Goldstone boson. In section \ref{5d} we turn to the 5d case. This case is a bit more subtle, as two possible theories, due to the orientifold action, are possible. After describing them, we turn to study flows in the singular and resolved spaces. A singularity in the background, already present in the trivial $AdS_6$ vacuum and with a clear string theory interpretation, plays a crucial role in selecting the dimensions of the operators taking VEV. In the broken baryon symmetry case we also identify the VEV of the condensate as well as the Goldstone boson. We finish in section \ref{conclusions} with some conclusions.

\section{4d $\mathcal{N}=2$ flows} \label{4d}

In this section we study RG flows in a 4d $\mathcal{N}=2$ gauge theory through its holographic dual. The simplest example is the so-called $A_1$ gauge theory, which can be engineered by placing $N$ D3 branes probing a $\mathbb{C}^2/\mathbb{Z}_2\times \mathbb{C}$ singularity \cite{Kachru:1998ys}. The CFT is a $SU(N)\times SU(N)$ gauge theory with global non-R symmetry $SU(2)_M\times U(1)_B$, being $SU(2)_M$ a mesonic global symmetry and $U(1)_B$ a baryonic symmetry. 

The types of flows which we will consider are triggered by motion on the moduli space, that is, by the VEV of certain operators. On general grounds we can imagine two types of such flows: one ``mesonic" type where all operators acquiring a VEV are neutral under the $U(1)_B$ and another ``baryonic" type where an operator charged under the $U(1)_B$ acquires a VEV. The former possibility corresponds to the case where the stack of D3 branes sourcing the geometry is located away from the tip of the cone, while the later possibility corresponds to the blow-up of the cone. We stress that in both cases the backgrounds corresponding to the flows asymptote to the same geometry, namely that of the singular cone. Hence both geometries indeed correspond to motion on the moduli space of the dual gauge theory --rather than turning on a deformation--.

In the gravity side, we consider a stack of D3 branes probing the --possibly resolved-- $\mathbb{C}^2/\mathbb{Z}_2\times \mathbb{C}$ geometry (we collect some useful details on the geometry of the $\mathbb{C}^2/\mathbb{Z}_2$ singularity in appendix \ref{geometry}). On general grounds we can write the standard Freund-Rubin ansatz for the background

\begin{equation}
ds^2=h^{-\frac{1}{2}}\,dx_{1,\,3}^2+h^{\frac{1}{2}}\,ds_6^2\qquad F_5=(1+\star)\,d(h^{-1})\wedge dx^0\wedge\cdots\wedge dx^3\, ;
\end{equation}
where $h$ is only a function of the internal coordinates and $ds_6^2$ is the internal space metric. The equation of motion of the 5-form field strength yields to

\begin{equation}
\label{eom4d}
d\star_6dh=\mathcal{C}\,\delta\, ,
\end{equation}
being $\star_6$ the Hodge dual with respect the metric in the internal space, $\mathcal{C}$ a normalization constant and $\delta$ the source term --in the end we have D3 branes somewhere in the cone--. In the following we will particularize this general equation to the cases of interest.

\subsection{Flows on the singular cone}\label{4dsingular}

We consider a stack of D3 branes at a certain point away from the tip of the singular $\mathbb{C}^2/\mathbb{Z}_2\times\mathbb{C}$ in the internal space.  The equation of motion (\ref{eom4d}) is just the Laplace equation in the internal space, which can be written as

\begin{equation}
\frac{1}{r_1^3}\,\partial_{r_1}\Big(r_1^3\,\partial_{r_1}h\Big)+\frac{4}{r_1^2}\,\Delta h+\frac{4}{r_1^2}\,\partial_{\psi}^2h+\frac{1}{r_2}\partial_{r_2}\Big(r_2\,\partial_{r_2}h\Big)+\frac{1}{r_2^2}\,\partial_{\chi}^2h=\frac{\mathcal{C}}{\sqrt{\rm{det}\,g_6}}\,\delta(X-X_0)\, ,
\end{equation}
where $X$ is a generic label for the internal coordinates and $X_0$ is the position where the stack is; and where $\mathcal{C}$ is a constant related to the $AdS_5$ radius $L$ as 

\begin{equation}
L^4=\frac{\mathcal{C}}{4\,{\rm vol}(S^5/\mathbb{Z}_2)}\, .
\end{equation}
Besides 

\begin{equation}
\Delta=\frac{1}{\sin\theta}\partial_{\theta}\Big(\sin\theta\,\partial_{\theta} \Big)+\Big(\frac{\partial_{\phi}}{\sin\theta}-\cot\theta\,\partial_{\psi}\Big)^2\, .
\end{equation}
Let us collectively denote the $\{\psi,\,\theta,\,\phi,\,\chi\}$ coordinates by $\xi$. Then the laplacian reads schematically $(\Delta_{r_1}+\Delta_{r_2}+\Delta_{\xi})\,h=\mathcal{C}\,\delta(r_1-r_1^0)\,\delta(r_2-r_2^0)\,\delta(\xi-\xi_0)$, being $\Delta_I$ the appropriate laplacians along the $X^I$ directions. 

Introduce now the functions $\tilde{Y}_{R,\,l,\,m}=e^{i\,m\,\phi}\,e^{i\,R\,\psi}\,J_{R,\,l,\,m}$ satisfying \cite{Benishti:2010jn}

\begin{equation}
\Delta\,\tilde{Y}_{R,\,l,\,m}=-\Big(l\,(l+1)-R^2\Big)\,\tilde{Y}_{R,\,l,\,m}\, .
\end{equation}
We can then construct

\begin{equation}
Y_I=e^{i\,n\,\chi}\,\tilde{Y}_{R,\,l,\,m}\, ,
\end{equation}
where $I$ collectively stands for all indices. Note that in order to have a well-defined solution, both $R\in\mathbb{Z}$ and $l\in\mathbb{Z}$, and so also $m\in \mathbb{Z}$. Besides, $n\in\mathbb{Z}$. Upon expanding 

\begin{equation}
h=\sum\,h_I(r_1,\,r_2)\,Y_I^{\star}(\xi_0)\,Y_I(\xi)\, ,
\end{equation}
then the equation of motion becomes of the form $\sum\,(\Delta_{r_1}+\Delta_{r_2}+f(r_1,\,r_2))\,h_I\,Y_I^{\star}(\xi_0)\,Y_I(\xi)=C\,\delta(r_1-r_1^0)\,\delta(r_2-r_2^0)\,\delta(\xi-\xi_0)$. Since $\tilde{Y}_{R,\,l,\,m}$ form a complete set of eigenfunctions, so do the $Y_I$. Hence $\sum_I\,Y_I^{\star}(\xi_0)\,Y_I(\xi)=\delta(\xi-\xi_0)$. Thus, in order to find a solution of the complete equation we need to demand $(\Delta_{r_1}+\Delta_{r_2}+f(r_1,\,r_2))\,h_I=\mathcal{C}\,\delta(r_1-r_1^0)\,\delta(r_2-r_2^0)$, which explicitly reads

\begin{equation}
\frac{1}{r_1^3}\,\partial_{r_1}\Big(r_1^3\,\partial_{r_1}h_I\Big)-\frac{4\,l(l+1)}{r_1^2}\, h_I+\frac{1}{r_2}\partial_{r_2}\Big(r_2\,\partial_{r_2}h_I\Big)-\,\frac{n^2}{r_2^2}\,h_I=\frac{\mathcal{C}}{r_1^3\,r_2}\,\delta(r_1-r_1^0)\,\delta(r_2-r_2^0)\, .
\end{equation}
In order to further proceed, it is useful to introduce polar coordinates as $r_1=\rho\,\cos\alpha$ and $r_2=\rho\,\sin\alpha$. In these coordinates the branes will be at $\{\rho_0,\,\alpha_0\}$. It is easy to check that the regular solutions are

\begin{equation}
\begin{array}{l c l}
h^{<}_{I}=\frac{\mathcal{C}}{4\,l+2\,n+4}\,\frac{1}{\rho_0^4}\,\Big(\frac{\rho}{\rho_0}\Big)^{2\,l+n}\,\cos^{2\,l}\alpha\,\sin^n\alpha & \leftrightarrow & \rho< \rho_0\, , \\
 &  &  \\
h^{>}_{I}=\frac{\mathcal{C}}{4\,l+2\,n+4}\,\frac{1}{\rho^{4}}\,\Big(\frac{\rho_0}{\rho}\Big)^{2\,l+n}\,\cos^{2\,l}\alpha\,\sin^n\alpha & \leftrightarrow & \rho> \rho_0 \, .
\end{array}
\end{equation}
Collecting all the pieces, we can write the warp factor in the $\rho>\rho_0$ region as

\begin{equation}
h=\frac{\mathcal{C}}{4\,\rho^4}+\sum_{l,\,n>0}\,\frac{\mathcal{C}}{4\,l+2\,n+4}\,\frac{1}{\rho^{4}}\,\Big(\frac{\rho_0}{\rho}\Big)^{2\,l+n}\,\mathcal{Y}_{I}(\alpha_0,\,\xi_0)^{\star} \,\mathcal{Y}_{I}(\alpha,\,\xi)\, ;
\end{equation}
while in the $\rho<\rho_0$ region it reads

\begin{equation}
h=\frac{\mathcal{C}}{4\,\rho_0^4}+\sum_{l,\,n>0}\,\frac{\mathcal{C}}{4\,l+2\,n+4}\,\frac{1}{\rho_0^{4}}\,\Big(\frac{\rho}{\rho_0}\Big)^{2\,l+n}\,\mathcal{Y}_{I}(\alpha_0,\,\xi_0)^{\star} \,\mathcal{Y}_{I}(\alpha,\,\xi)\, ;
\end{equation}

being

\begin{equation}
\mathcal{Y}_{I}(\alpha,\,\xi)=\cos^{2\,l}\alpha\,\sin^n\alpha\,Y_{I}(\psi,\,\theta,\,\phi,\,\chi)\, .
\end{equation}
Note in particular that, as advertised above, the geometry asymptotes just like the singular cone. Indeed $h$, in the $\rho>\rho_0$ region, starts as $\rho^{-4}$.

\subsubsection{An alternative computation}\label{4dsingularalternative}

The coordinates used in the previous section are adapted to the orbifold, as the $A_1$ singularity is explicitly separated from the $\mathbb{C}$ factor. However, upon a change of coordinates, the metric in the internal space can be written as

\begin{equation}
ds_6^2=d\rho^2+\rho^2\,ds_{S^5/\mathbb{Z}_2}^2\, .
\end{equation}
The equation for $h$ is just

\begin{equation}
\frac{1}{\rho^5}\,\partial_{\rho}\Big(\rho^5\,\partial_{\rho}\,h\Big)+\frac{1}{\rho^2}\,\Delta_{S^5/\mathbb{Z}_2}\,h=\mathcal{C}\,\delta\, ,
\end{equation}
being $\Delta_{S^5/\mathbb{Z}_2}$ the laplacian on the $S^5/\mathbb{Z}_2$. Since that space is locally identical to $S^5$, the equation of motion looks exactly like the $S^5$ one. Denoting the $S^5$ angular coordinates by $\bar{\xi}$ and setting

\begin{equation}
h=\sum\,h_I(\rho)\,\mathcal{Y}_I^{\star}(\bar{\xi}_0)\,\mathcal{Y}_I(\bar{\xi})\, ,
\end{equation}
with $\mathcal{Y}_I$ the $S^5$ spherical harmonics, the $h_I(\rho)$ eom is

\begin{equation}
\frac{1}{\rho^5}\,\partial_{\rho}\Big(\rho^5\,\partial_{\rho}\,h_I\Big)-\frac{\ell\,(\ell+4)}{\rho^2}\,h_I=\frac{\mathcal{C}}{\rho^5}\,\delta(\rho-\rho_0)\, ;
\end{equation}
whose solution is

\begin{equation}
\begin{array}{l c l}
h^{<}_{I}=\frac{\mathcal{C}}{4+2\,\ell}\, \frac{1}{\rho_0^4}\,\Big(\frac{\rho}{\rho_0}\Big)^{\ell} & \leftrightarrow & \rho< \rho_0\, , \\
 &  &  \\
h^{>}_{I}=\frac{\mathcal{C}}{4+2\,\ell}\,\frac{1}{\rho^4}\,\Big(\frac{\rho_0}{\rho}\Big)^{\ell} & \leftrightarrow & \rho> \rho_0 \, .
\end{array}
\end{equation}
We have not yet taken into account the orbifold. Prior to orbifolding, we see that modes are classified into representations of spin $\ell$ of $SO(6)$. Decomposing such representations into $U(1)_{\chi}\times SU(2)_M\times SU(2)_R$, the orbifold selects a subset of representations with even spin, that is $\ell=2\,l+n$, where $2\,l$ stands for the orbifold, hence recovering the results in the previous analysis.

\subsection{Flows on the resolved cone}\label{4dresolved}

We now consider the resolution of the cone (see appendix \ref{geometry}). Then eq.(\ref{eom4d}) becomes 

\begin{equation}
r_1^{-3}\,\partial_{r_1}\Big(r_1^3\,f\,\partial_{r_1}h\Big)+\frac{4}{r_1^2}\,\Delta h+\frac{4}{r_1^2\,f}\,\partial_{\psi}^2h+\frac{1}{r_2}\partial_{r_2}\Big(r_2\,\partial_{r_2}h\Big)+\frac{1}{r_2^2}\,\partial_{\chi}^2h=\mathcal{C}\,\delta\, ,
\end{equation}
where we collectively denote the sources by $\delta$ and where

\begin{equation}
\Delta=\frac{1}{\sin\theta}\partial_{\theta}\Big(\sin\theta\,\partial_{\theta} \Big)+\Big(\frac{\partial_{\phi}}{\sin\theta}-\cot\theta\,\partial_{\psi}\Big)^2\, .
\end{equation}

Just as in the singular case we can write

\begin{equation}
h=\sum_{I}\,h_{I}(r_1,\,r_2)\,Y_{I}^{\star}(\xi_0)\,Y_{I}(\xi)\, .
\end{equation}
Hence, the equation for $h_I$ (away from the sources) is

\begin{equation}
r_1^{-3}\,\partial_{r_1}\Big(r_1^3\,f\,\partial_{r_1}h_{I}\Big)-\frac{4\,\Big( l(l+1)-R^2\Big)}{r_1^2}\, h_{I}-\frac{4\,R^2}{r_1^2\,f}\,h_{I}+\frac{1}{r_2}\partial_{r_2}\Big(r_2\,\partial_{r_2}h_{I}\Big)-\,\frac{n^2}{r_2^2}\,h_{I}=0\, .
\end{equation}
Since the D3 branes will be located at $r_1=c$ and $ r_2=0$, where the $\chi$ circle and the $\psi$ circle shrink, it's reasonable to truncate to $R=n=0$. Then, the equation to solve is

\begin{equation}
\frac{1}{r_1^3}\,\partial_{r_1}\Big(r_1^3\,f\,\partial_{r_1}h_{I}\Big)-\frac{4\, l(l+1)}{r_1^2}\, h_{I}+\frac{1}{r_2}\partial_{r_2}\Big(r_2\,\partial_{r_2}h_{I}\Big)=0\, .
\end{equation}

Let us now switch to the $\{\rho,\,\alpha\}$ coordinates. Unfortunately one finds a fairly complicated equation which we have not been able to solve exactly. Nevertheless, we can study its large-$\rho$ asymptotic properties. To that matter, we set to first order

\begin{equation}
h_I=\Big(\frac{c}{\rho}\Big)^a\,f(\alpha)\, ,
\end{equation}
and expand in powers of $\frac{c}{\rho}$. The leading term determines $f(\alpha)$ as

\begin{equation}
\left(-4 a+a^2-2 m^2+(-4+a)\, a \,\cos(2 \alpha )\right)\, f\,\sin\alpha +2 \left(\cos(3 \alpha )\, f'+\cos^2\alpha \sin \alpha \, f''\right)=0\, ;
\end{equation}
where for simplicity we have set $m^2=4\,l\,(l+1)$. The solution to this is

\begin{equation}
f=\cos^{-1+\sqrt{1+m^2}}\alpha\,\,_2F^1[\frac{3}{2}-\frac{a}{2}+\frac{\sqrt{1+m^2}}{2},\,-\frac{1}{2}+\frac{a}{2}+\frac{\sqrt{1+m^2}}{2},\,1+\sqrt{1+m^2},\,\cos^2\alpha]\, .
\end{equation}
Regularity at $\alpha=0$ demands

\begin{equation}
a=4+\sqrt{1+m^2}-1+2\,q=4+2\,l+2\,q\, ,
\end{equation}
for $q\in\mathbb{Z}$. This integer arises since regularity at $\alpha=0$ demands a certain $\Gamma(x)^{-1}$ function to vanish, which happens for $x=-q$. This integer should not be confused with the $n$ in section \ref{4dsingular}, as the later is related to the $U(1)$ charge conjugate to $\chi$, while $q$ is not related to any charge.

We can  directly read off the dimension of the modes from here, which is just $a-4$. Hence $\Delta=2\,l+2\,q$. Note that, since $a=4+\Delta$, we again have that, as promised, the geometry asymptotically becomes just the same as the singular cone. This explicitly reflects the fact that also the blow-up geometry corresponds to a flow triggered not by a deformation of the gauge theory but by a VEV.

\subsection{Gauge theory}\label{GT4d}

The $AdS/CFT$ duality implies that the above geometries describe two different RG flows along which certain operators $\mathcal{O}_I$ acquire a VEV. Such operators correspond to the wavefunctions

\begin{equation}
\langle\mathcal{O}_I\rangle=\rho_0^{2\,l+m\,n}\,\mathcal{Y}_{I}(\xi_0)^{\star}\, ,
\end{equation}
being $m=1$ for the singular case and $m=2$ for the resolved case. 

In order to identify the gauge theory operators corresponding to the $\mathcal{Y}_{I}(\xi_0)^{\star}$, let us first turn to the algebraic geometry of the $\mathbb{C}^2/\mathbb{Z}_2\times \mathbb{C}$. Introducing complex coordinates $z_i$ related to the real ones as

\begin{equation}
z_1=\rho\,\cos\alpha\,e^{i\frac{\psi+\phi}{2}}\,\cos\frac{\theta}{2}\, ,\qquad z_2=\rho\,\cos\alpha\,e^{i\frac{\psi-\phi}{2}}\,\sin\frac{\theta}{2}\, ,\qquad z_3=\rho\,\sin\alpha\,e^{i\chi}\, ,
\end{equation}
we have that the $\mathbb{C}^2/\mathbb{Z}_2\times \mathbb{C}$ space is defined by the $z_i$ coordinates together with the identification $(z_1,\,z_2)\sim(-z_1,\,-z_2)$. 

Let us strip off the $\mathbb{C}$ factor and concentrate on the $\mathbb{C}^2/\mathbb{Z}_2$ space. In complex coordinates, the metric is simply $ds_{A_1}^2=dz_1\,d\bar{z}_1+dz_2\,d\bar{z}_2$. This is invariant under $SU(2)_M\times SU(2)_R$. The $SU(2)_M$ symmetry acts on the $(z_1,\,z_2)$ doublet, while the $SU(2)_R$ symmetry acts on the $(z_1,\,\bar{z}_2)$ doublet. Both these actions are compatible with the orbifold projection for this $\mathbb{Z}_2$ case --since it acts diagonally by a $-1$ on either doublet--. 

Let us first forget about the orbifold projection. Out of the $SU(2)_R$, only its $U(1)_R$ subgroup is manifest. Hence, it is useful to consider the manifest symmetry subgroup $SU(2)_M\times U(1)_R$. The charges of the $\{z_i,\,\bar{z}_i\}$ under its $U(1)_M\times U(1)_R$ cartan are

\begin{equation}
\begin{array}{c | c c c c}
 & z_1 & z_2 & \bar{z}_1 & \bar{z}_2 \\ \hline
 U(1)_M & \frac{1}{2} & -\frac{1}{2} & -\frac{1}{2} & \frac{1}{2} \\
 U(1)_R & 1 & 1 & -1 & -1
 \end{array}
 \end{equation}
 Following \cite{Klebanov:2007us}, we consider forming the states
 
 \begin{equation}
 \label{states0}
 |\frac{n}{2},\,m\rangle=\sqrt{\frac{1}{(\frac{n}{2}+m)!\,(\frac{n}{2}-m)!}}\,z_1^{\frac{n}{2}+m}\,z_2^{\frac{n}{2}-m}\, ,\qquad  \overline{|\frac{n}{2},\,m\rangle}=\sqrt{\frac{1}{(\frac{n}{2}+m)!\,(\frac{n}{2}-m)!}}\,\bar{z}_2^{\frac{n}{2}+m}\,\bar{z}_1^{\frac{n}{2}-m}\, .
 \end{equation}
Here $m\in[-\frac{n}{2},\,\frac{n}{2}]$. Such states have the following charges

\begin{equation}
\begin{array}{c | c  c}
 & |\frac{n}{2},\,m\rangle & \overline{|\frac{n}{2},\,m\rangle}  \\ \hline
 U(1)_M & m & m \\
 U(1)_R & n & -n
 \end{array}
 \end{equation}
Thus these states are the states in a spin $l=\frac{n}{2}$ representation of $SU(2)$ --hence the labels of the states--. On the other hand, for the $ |\frac{n}{2},\,m\rangle$, since $l=\frac{n}{2}$ and $R=n$, it is clear that $\frac{n}{2}=\frac{l}{2}+\frac{R}{4}$, while for the $\overline{|\frac{n}{2},\,m\rangle}$, since $l=\frac{n}{2}$ and $R=-n$, $\frac{n}{2}=\frac{l}{2}-\frac{R}{4}$. Hence, we can label the states as

\begin{equation}
\label{Ytilde}
 |\frac{n}{2},\,m\rangle=|\frac{l}{2}+\frac{R}{4},\,m\rangle\, ,\qquad \overline{|\frac{n}{2},\,m\rangle}=|\frac{l}{2}-\frac{R}{4},\,m\rangle\, .
 \end{equation}
With these we can now construct arbitrary $SU(2)_M$ and $U(1)_R$ states. These are labelled by the $SU(2)_M$ quantum numbers $\{l,\,m\}$ and by the $R$ charge as $|l,\,m;\,R\rangle$, and are constructed as

\begin{equation}
 \label{states}
|l,\,m;\,R\rangle=\sum\,C_{\{\{\frac{l}{2}+\frac{R}{4},\,m_1\},\,\{\frac{l}{2}-\frac{R}{4},\,m_2\}\}}^{\{l,\,m\}}\,|\frac{l}{2}+\frac{R}{4},\,m_1\rangle\,|\frac{l}{2}-\frac{R}{4},\,m_2\rangle\, .
\end{equation}
where $C_{\{\{j_1,\,m_1\},\,\{j_2,\,m_2\}\}}^{\{j,\,m\}}$ are the top spin Clebsch-Gordon coefficients. One can then easily check that the states $|l,\,m;\,R\rangle$, when written in terms of the real coordinates, become just the $\tilde{Y}_{R,\,l,\,m}$ functions.

We are still not done, as we need to recover the $\mathbb{C}$ factor. It is parametrized by the complex coordinate $z_3$, and the metric is $ds^2_{\mathbb{C}}=dz_3\,d\bar{z}_3$, which is obviously invariant under a $U(1)_{\chi}$ symmetry. The relevant wavefunctions are simply, up to a normalization, $|n\rangle=z_3^n$, which have $U(1)_{\chi}$ charge $n$. It then follows that 

\begin{equation}
\mathcal{Y}_I=|n\rangle\,|l,\,m;\,R\rangle\, .
\end{equation}
Note that $z_3^n$ adds a factor of $\sin^n\alpha$, so that the $\mathcal{Y}_I$ recover the expression of the wavefunctions obtained from the gravity side.

We still need to consider the effect of the orbifold projection. Upon acting with the orbifold, the states in eq.(\ref{states0}) pick a factor $(-1)^{\pm n}$. Hence, the surviving states are those for which $n$ is even. This translates into the fact that the only allowed states are those with even $R$.

We now turn to the gauge theory. It is a $\mathcal{N}=2$ $SU(N)\times SU(N)$ gauge theory with 2 hypers which can be broken in $\mathcal{N}=1$ language into two bifundamentals in the $(\Box,\,\bar{\Box})$ and two bifundamentals in the $(\bar{\Box},\,\Box)$. The $W$ is

\begin{equation}
W=\phi_1\,A_i\,B_j\,\epsilon^{ij}+\phi_2\,B_i\,A_j\,\epsilon^{ij}\, .
\end{equation}
The theory shows an $SU(2)_M$ symmetry rotating $A_i$ and $B_i$. Besides, it has a $U(1)_r\times SU(2)_R$ R-symmetry. The cartan of the non-abelian factor of the R-symmetry will be denoted $U(1)_R$, while the cartan of the $SU(2)_M$ will be denoted by $U(1)_M$. The charges of each field under such cartans are

\begin{equation}
\begin{array}{c | c c c c c c}
 & \phi_1 & \phi_2 & A_1 & A_2 & B_1 & B_2 \\ \hline
 U(1)_M & 0 & 0 & \frac{1}{2} & -\frac{1}{2}& \frac{1}{2}& -\frac{1}{2}\\
 U(1)_R & 0 & 0 &1 & 1 & 1 &1 \\
  U(1)_r & \frac{2}{3} & \frac{2}{3} & \frac{2}{3}& \frac{2}{3} & \frac{2}{3}& \frac{2}{3} 
 \end{array}
 \label{charges}
 \end{equation}
Note that we can combine $U(1)_R$ and $U(1)_r$ into a new $U(1)_{\chi}$ such that only the $\phi_i$ are charged.

The F-terms are

\begin{equation}
A_1\,B_2=A_2\,B_1\, ,\qquad B_1\,A_2=B_2\,A_1\, ,\qquad B_i\,\phi_1=\phi_2\,B_i\, ,\qquad \phi_1\,A_i=A_i\,\phi_2\, .
\end{equation}
Let us start considering operators purely in the Higgs branch, \textit{i.e.} those made out of $A_i$ and $B_j$ with no adjoints. Imposing the F-terms, we can construct the following three operators $u=A_1\,B_1$, $v=A_2\,B_2$ and $w=A_1\,B_2$ subject to the relation $u\,v=w^2$ defining the $\mathbb{C}^2/\mathbb{Z}_2$ singularity. We can solve this relation introducing $u=z_1^2$, $v=z_2^2$ and $w=z_1\,z_2$. Clearly $(z_1,\,z_2)\sim-(z_1,\,z_2)$. Hence this way we have an explicit mapping between the states constructed above and the operators in field theory, which we will collectively denote by $\mathcal{O}_I$.

Denoting by $J_3^{M,\,R}$ the third component of the $SU(2)_{M,\,R}$, one can check that the states with the maximal $J_3^{M,\,R}$ are of the form $z_1^{2\,l}$, which becomes $(A_1\,B_1)^{2\,l}$. Lower $J_3^{M,\,R}$ operators generically involve both $z_i$ and $\bar{z}_i$, and hence correspond to non-holomorphic operators. However, they can be thought as descendants obtained by repeatedly acting with $SU(2)_{M,\,R}$ raising/lowering operators, acting on the doublets $(A_i,\,B_i)$ and $(A_i,\,\epsilon_{ij}\,B_j^{\dagger})$ respectively. Hence, every operator can be regarded as belonging to the same multiplet as a purely holomorphic operator, hence having the same dimension equal to the classical one.

Let's now look at operators in the Coulomb branch. These will involve insertions of $\phi_i^n$, and hence will have $n$ charge under $U(1)_{\chi}$. Note also that the theory has a $\mathbb{Z}_2$ symmetry exchanging the two gauge groups while at the same time doing

\begin{equation}
\phi_1\leftrightarrow \phi_2\, ,\qquad A_i\leftrightarrow B_i\, .
\end{equation}
This is just the quantum symmetry of the orbifold. We can construct operators which are not invariant under the symmetry. Consider for example

\begin{equation}
\mathcal{O}_+ = {\rm Tr}\,\phi_1^2+{\rm Tr}\,\phi_2^2\, ,\qquad \mathcal{O}_-={\rm Tr}\,\phi_1^2-{\rm Tr}\,\phi_2^2\, .
\end{equation}
Clearly, only $\mathcal{O}_+$ is invariant, while $\mathcal{O}_-$ picks a minus sign. As only twisted sector states are charged under the quantum symmetry of the orbifold, we must conclude that the $\mathcal{O}_-$ field comes from the twisted sector. On the other hand, it is natural to expect that the operators capturing the motion of the branes are neutral under this symmetry. This naturally leads to consider the branch where $\phi_1=\phi_2$, so that $\phi^n$ naturally corresponds to the $z_3^n$ above. In fact, the $U(1)_{\chi}$ charge is precisely the expected $n$.

\subsubsection{The singular case}

Since in the singular flows we placed the branes away from the tip of the singular cone, it is natural to expect that no twisted sector field will get a VEV. Hence, let us consider the branch of the moduli space which freezes  twisted sector fields by setting $\phi=\phi_1=\phi_2$. Then, the F-term relations imply that $\phi_{1,\,2}$ commute with $A_i$ and $B_i$. Therefore a generic operator can be written as ${\rm Tr}\,\phi^n\,\mathcal{O}_I$. This exactly mimics the structure of the wavefunctions which we have found, as indeed the $|n\rangle$ state corresponds to the $\phi^n$. Such operators come in spin $l$ representations of $SU(2)_M$ and have dimension $2\,l+n$, exactly as the modes found in supergravity. Thus, since only mesonic fields take a VEV, the dual gauge theory is in a mesonic branch.

Note that we could place the stack of branes at the origin of the $\mathbb{C}$ plane, which would force us to impose $n=0$. Then the operators taking a VEV are purely mesonic operators on the Higgs branch, which is then isomorphic to $\mathbb{C}^2/\mathbb{Z}_2$. Alternatively we could place the branes at the origin of the $\mathbb{C}^2/\mathbb{Z}_2$, which would demand to set $l=0$. Hence the operators taking a VEV would be those of the form $\phi^n$.

\subsubsection{The resolved case and baryonic symmetry breaking}

Recall that in this case we truncated $R=0$ and $n=0$. Hence the wavefunctions for the modes can be translated into operators following the general discussion above upon setting $R=0$ and $n=0$. The dimension of these operators is $\Delta=2\,l+2\,q$. Setting for the moment $q=0$, we see that we get a tower of operators with spin $l$ under $SU(2)_M$ and neutral under $SU(2)_R$ --because $R=0$--. The lowest operator is $l=1$, which has dimension 2 and includes, as its $m=0$ component, the familiar  $\mathcal{U}\sim A_i^{\dagger}\,A_i-B_i\,B_i^{\dagger}+[\phi_1,\,\bar{\phi_1}]-[\phi_2,\,\bar{\phi}_2]$ operator, which is part of the baryonic current multiplet, just as in \cite{Klebanov:2007us}, thus supporting the claim that the resolved background corresponds to a spontaneously broken baryonic symmetry phase.

For higher $l$ modes one can simply use the dictionary in our general discussion above and find the corresponding operator $\mathcal{O}_I$. In addition, analogously to the conifold case \cite{Klebanov:2007us}, one can check that going one order further in $\frac{c}{\rho}$, the dimension of the operator is $\Delta+2$, which suggests that these higher order terms corresponds to instertions of powers of $\mathcal{U}$, that is, tooperators of the form $\mathcal{U}^n\,\mathcal{O}_I$ for $n=1,\,2,\cdots$.

Let us now turn to the integer $q$. Again the wavefunctions can be read off from our general discussion just by setting $R=0$ and $n=0$. It thus seems that there is no room for the extra integer $q$. Nevertheless one can imagine constructing an operator of the form $\mathcal{T}\sim \phi_1\,\phi_1^{\dagger}-\phi_2\,\phi_2^{\dagger}$ which is only charged under $U(1)_r$ and has dimension 2, at least classically. Even though this is a non-chiral field, based on the findings on the holographic dual, we conjecture that the classical dimension continues to hold, so that the modes with higher $q$ correspond to insertions of this operator, that is, to $\mathcal{T}^q\,\mathcal{O}_I$. Note that this is a twisted sector field, which fits naturally within the blow-up scenario. 

As we have argued, among the lowest dimension operators taking a VEV we find the scalars in the conserved current multiplet. Hence it is natural to identify this background with a phase of the gauge theory where the baryonic symmetry has been spontaneously broken by giving a VEV to a single chiral field, which, with no loss of generality we can take as $A_1= c\,\unity$. This gives a VEV to a dimension $\Delta=N$ baryonic operator $\mathcal{B}_A=A_1^N\sim c^N$. Such VEV is captured holographically by the so-called baryonic condensate, namely an euclidean D3 brane wrapping $X=\{r_1,\,r_2,\,\psi,\,\chi\}$. Its DBI action is

\begin{equation}
S_{DBI}=\frac{T_3}{2}\,\int_X\, r_1\,r_2\,h\, .
\end{equation}
Plugging in the expression for the warp factor we find\footnote{Note that one can re-write this warped volume in terms of the K\"{a}hler form of the cone, finding $S_{DBI}=\frac{T_3}{4}\,\int_X\,h\,\frac{\omega^2}{2!}$, and hence adapt the proof in \cite{Benishti:2010jn} to the $CY_3$ case to conclude that VEV of the baryon condensate will have the expected form.}

\begin{equation}
S_{DBI}=\frac{T_3}{2}\,(2\,\pi)^2\,\sum_l \tilde{Y}_{0,\,l,\,0}^{\star}(\xi_0)\,\tilde{Y}_{0,\,l,\,0}(\xi)\,\int \, r_1\,r_2\,h_l(r_1,\,r_2)\, .
\end{equation}
The integrand scales asymptotically as $\rho^{-1-2\,l}$. Hence all the $l\ne 0$ terms will give some finite $\rho$ integral, while the $l=0$ will give a logarithmically divergent integral which has to be cut-off at some $\rho_c$. So we can separate the $l=0$ term and write

\begin{equation}
S_{DBI}=\frac{T_3}{2}\,(2\,\pi)^2\,\int \, r_1\,r_2\,h_0(r_1,\,r_2)+\frac{T_3}{2}\,(2\,\pi)^2\,\sum_{l>0} \tilde{Y}_{0,\,l,\,0}^{\star}(\xi_0)\,\tilde{Y}_{0,\,l,\,0}(\xi)\,\int \,r_1\,r_2\,h_l(r_1,\,r_2)\, .
\end{equation}
It is easy to show that the divergent term diverges like $S_{DBI}=N\,\log\rho_c$, being $\rho_c$ a UV cut-off. This corresponds to the VEV of dimension $\Delta=N$ operator, as expected for a baryon VEV. 

Alternatively, we can consider a $D3$ brane wrapping the blown-up $S^2$. Its action is simply

\begin{equation}
S_{DBI}=-T_3\,c^2\,\pi\,\int \,dx^0\,dx^1\, .
\end{equation}
This finite-tension object, which looks like a cosmic string from the field theory perspective and is ``electric-magnetic" dual to the baryon condensate, would source a $\delta C_4$ such that

\begin{equation}
\delta F_5=(1+\star)da_2\wedge W\, ,
\end{equation}
being $W$ a 2-form in the internal space which, for $r\sim c$, becomes the volume form of the blown-up cycle while $a_2$ is a 2-form in the Minkowski directions whose existence, along the lines in \cite{Klebanov:2007cx}, we assume here. The $\delta F_5$ then contains a piece with $\delta F_5\supset\star_4 da_2\wedge (h\,\star_6\,W)$. We can locally write $\star_4da_2=dp$, so a local integration is $\delta C_4=p\,(h\,\star_6\,W)$. Note that $\star_6W$ is a 4-form transverse, in the internal space, to the blown-up 2-cycle, that is, precisely along the directions wrapped by the baryonic condensate brane. Thus, its WZ action is of the form $e^{i\,T_3\,p\,\int h\,\star_6\,W}$, that is, becomes the phase of the VEV of the baryon. Then the Minkowski scalar $p$ is naturally identified with the Goldstone boson of the broken global symmetry. Furthermore, the equations of motion of $\delta F_5$ demand that $p$ satisfies $d\star_4dp=\delta(x-x_0)$ begin $x_0$ the position of the cosmic string --D3 brane-- in the Minkowski, thus showing how the Goldstone boson is an axionic field winding around the defect arising from spontaneous symmetry breaking \cite{Klebanov:2007cx}.

\section{5d RG flows}\label{5d}

We now turn to the 5d case. Naively, gauge theories in 5d are non-renormalizable. Nevertheless it was shown in \cite{Seiberg:1996bd} that, under certain circumstances, upon appropriately choosing gauge group and matter content they can be at fixed points. As these theories admit a large N limit   \cite{Intriligator:1997pq}, it is natural to look for a gravity dual. 

For the class of theories discussed in \cite{Seiberg:1996bd} the gravity dual was found in \cite{Brandhuber:1999np} by considering $N_f$ D8 branes on top of an $O8^-$ probed by $N$ D4 branes. If $N_f<8$, and upon tuning the dilaton to diverge on top of the 8-brane stack, the near brane geometry becomes $AdS_6\times \hat{S}^4$, being $\hat{S}^4$ half of a 4-sphere due to the left-over of the orientifold projection in the near brane region. On the other hand, the dual field theory is a $USp(2\,N)$ gauge theory with one antisymmetric hypermultiplet. 

As opposed to lower dimensionalities, $AdS_6$ geometries are very scarce \cite{Passias:2012vp}. \footnote{See however \cite{Lozano:2012au, Lozano:2013oma} for new $AdS_6$ geometries.} However, the basic $AdS_6/CFT_5$ duality can be naturally extended by orbifolding the internal space \cite{Bergman:2012kr}. The geometry becomes $AdS_6\times \hat{S}_4/\mathbb{Z}_k$, while  the dual fixed point theory is a quiver gauge theory whose details depend on whether $k$ is even or odd and, in the former case, how exactly the orientifold projection acts.

In order to ease the discussion, let us concentrate from now on on the case of a $\mathbb{Z}_2$ orbifold and set, for simplicity, $N_f=0$. As discussed in \cite{Bergman:2012kr}, there are two possible orientifold projections, depending on their action on the twisted sector of the orbifold projection. In the so-called vector structure (VS) case the orientifold projection keeps a 5d hypermultiplet from the orbifold twisted sector so that the resulting theory is a $USp(2\,N)\times USp(2\,N)$ gauge theory with bifundamental matter. In turn, the so-called no-vector structure (NVS) projection keeps a 5d vector multiplet and leads to a $SU(2\,N)$ gauge theory with 2 antisymmetric hypermultiplets. 

Both theories correspond to $N$ D4 branes probing an $O8^-$ wrapping $\mathbb{C}^2/\mathbb{Z}_2$ and are dual to exactly the same $AdS_6\times \hat{S}^4/\mathbb{Z}_2$ background. Reassuringly, in both cases the classical flavor-blind sector of the Higgs branch --probed by dual giant gravitons spinning in $\hat{S}^4/\mathbb{Z}_2$-- is precisely the $A_1$ singularity \cite{Bergman:2012qh} (see also \cite{Bergman:2013koa} for further holographic checks regarding operator counting in orbifold theories).

We are interested in exploring flows in these theories triggered by VEVs of operators. Just as in the 4d case discussed in the previous section, in the gravity dual we can imagine either moving the stack of branes sourcing the geometry away from the orbifold singularity or alternatively  blowing up the singularity. Note that, on general grounds, if the singularity is blown-up a brane behaving as a cosmic string will be allowed \cite{Bergman:2012kr}. Hence exactly as in the 4d case it is natural to expect the flows in the singular cone to be triggered by mesonic VEV's while those on the resolved cone to be triggered by baryonic VEV's. However, while the NVS has a $U(1)_B$ baryonic symmetry the VS case does not. This is inherited from the orientifold projection, which in the VS case removes the blow-up mode for the orbifold singularity while keeps it in the NVS case. Hence, even though the same $AdS_6\times \hat{S}_4$ background is dual to both the VS and the NVS theories, the spectrum of allowed fluctuations will be different. In particular the resolution mode is only present in the NVS case. Thus, when dealing with flows in the resolved conifold, we must refere to the NVS case, where the blow-up mode is allowed and the dual CFT has a $U(1)_B$ symmetry.

On general grounds, the geometry sourced by $N$ D4 branes probing a stack of $N_f$ D8 branes on top of an $O8^-$ plane must be of the form

\begin{equation}
ds^2=\Omega^2(z)\Big\{ h^{-\frac{1}{2}}\,dx_{1,\,4}^2+h^{\frac{1}{2}}\,\Big[ds_4^2+dz^2\Big]\Big\}\, ,
\end{equation}
where

\begin{equation}
e^{-\Phi}=H^{\frac{5}{4}}\,h^{\frac{1}{4}}\, ,\qquad \Omega=\Big(\frac{3}{2}\,m\,z\Big)^{-\frac{1}{6}}\, ,\qquad H = \Big(\frac{3}{2}\,m\,z\Big)^{\frac{2}{3}}\, .
\end{equation}
In addition to the Romans mass $F_0=m$, there is a 6-form field strength  given by the standard Freund-Rubin ansatz

\begin{equation}
F_6=d\,( h^{-1})\,dx_0\wedge dx_1\wedge dx_2\wedge dx_3\wedge dx_4\, .
\end{equation}
The equations of motion of the 6-form fix $h$ to satisfy

\begin{equation}
d(\Omega^{-2}\,\star_5\,dh)=\mathcal{C}\,\delta\, ;
\end{equation}
where $\star_5$ is the Hodge-star operator with respect to the $ds_4^2+dz^2$ metric, $\mathcal{C}$ a normalization constant proportional to $N$ and $\delta$ a source term supported at the location of the D4 branes. 

The $ds_4^2$ stands for the metric on the transverse space to the D4's inside the 8-branes. In the singular case it is just the singular $A_1$ space. Locating the D4's at the tip we obtain the warped $AdS_6\times \hat{S}^4/\mathbb{Z}_2$ geometries of \cite{Bergman:2012kr}. The warp factor of the $AdS_6$ is a function of the azimuthal angle $\alpha$ of the $\hat{S}^4$. Moreover, one can see that both curvature and dilaton diverge at the north pole of the $\hat{S}^4$ when $\alpha=0$, which is then identified with the location of the orientifold. This should expected since, in order to find the fixed point theory, we need to remove the bare YM coupling, which in the gravity side amounts to tune the dilaton to diverge on top of the orientifold plane. \footnote{In fact, due to string duality, this suggests the emergence of enhanced global symmetries in these fixed point theories \cite{Seiberg:1996bd}.} Note that, even though the geometry blows-up at $\alpha=0$ and in principle the full string theory should be used to describe the physics there, at least certain quantities are well-defined in the supergravity solution. In fact, the part of the Higgs branch not involving fundamental fields is well-captured by giant gravitons spinning at $\alpha=0$ \cite{Bergman:2012qh}. 

In the following we will be interested on moving the branes away from the singularity as well as on replacing the singular $A_1$ space by its resolution. On general grounds we expect the first type of geometries to correspond to RG flows triggered by mesonic VEV's --and hence possible in both the VS and NVS cases-- while the second ones to correspond to flows triggered by baryonic VEV's. Of course, as described above, the latter can only happen in the NVS case.

\subsection{Flows on the singular space}

Let us concentrate on flows on the singular space. The equation of motion for $h$ is

\begin{equation}
\frac{1}{r^3}\, \partial_{r}\Big(r^3 \,\partial_r h \Big)+\frac{1}{z^{\frac{1}{3}}}\,\partial_{z}\Big(z^{\frac{1}{3}}\,\partial_z h \Big)+ \frac{4}{r^2}\,\Delta\,h+\frac{4}{r^2}\,\partial_{\psi}^2\,h=\mathcal{C}\,\delta\, .
\end{equation}
We localize the branes at a given point $\xi_0$ in the angular coordinates $\xi=\{\psi,\,\theta,\,\phi\}$. Following the same procedure as above we expand

\begin{equation}
h=\sum\,h_I(r,\,z)\,\tilde{Y}^{\star}_I(\xi_0)\,\tilde{Y}_I(\xi)\, ,
\end{equation}
where the $\tilde{Y}_I$ functions are the eigenfunctions of the $\Delta$ laplacian introduced above. Note that now we do not have the circle $\chi$ --the transverse space to the $A_1$ singularity in the internal directions is just a line--. Hence our multi-index $I$ only involves $\{R,\,l,\,m\}$. We find the following equation for the function $h_I$

\begin{equation}
\frac{1}{r^3}\, \partial_{r}\Big(r^3 \,\partial_r h_I \Big)+\frac{1}{z^{\frac{1}{3}}}\,\partial_{z}\Big(z^{\frac{1}{3}}\,\partial_z h_I \Big)- \frac{4\,l\,(l+1)}{r^2}\,h_I=\mathcal{C}\,\delta\, .
\end{equation}
It is useful to switch to polar coordinates in the $\{r,\,z\}$ plane defining $r=\rho\,\cos\alpha$ and $z=\rho\,\sin\alpha$. We then find

\begin{equation}
\begin{array}{l c l}
h^{<}_{I}=\Big(\frac{\rho}{\rho_0}\Big)^{a}\,\cos^{2\,l}\alpha\,\, _2F^1[-\frac{a}{2}+l,\,\frac{5}{3}+\frac{a}{2}+l,\,2+2l,\,\cos^2\alpha] & \leftrightarrow & \rho< \rho_0\, , \\
 &  &  \\
h^{>}_{I}=\frac{1}{\rho^{\frac{10}{3}\,}}\,\Big(\frac{\rho_0}{\rho}\Big)^{a}\,\cos^{2\,l}\alpha\,\, _2F^1[-\frac{a}{2}+l,\,\frac{5}{3}+\frac{a}{2}+l,\,2+2\,l,\,\cos^2\alpha]  & \leftrightarrow & \rho> \rho_0 \, .
\end{array}
\end{equation}
As described above, the north pole of the $S^4$ has to be treated with care in this case, as both dilaton and curvature diverge there. Hence, we need to carefully discuss the boundary conditions to be imposed on $\alpha$. To that matter, let us consider a generalized version of the eom in the singular case

\begin{equation}
\frac{1}{r_1^3}\, \partial_{r_1}\Big(r_1^3 \,\partial_{r_1}\varphi\Big)+\frac{1}{r_2^{a}}\,\partial_{r_2}\Big(r_2^{a}\,\partial_{r_2} \varphi \Big)- \frac{4\,l\,(l+1)}{r_1^2}\,\varphi=0\, ,
\end{equation}
so that $a=1$ is the eom for the $AdS_5$ case and $a=\frac{1}{3}$ the eom for the $AdS_6$ case. 

The equation depends on the eigenvalue $l$. We can consider two copies of the equation for modes for different eigenvalues $l_{1,\,2}$. Manipulating them, we find
 
 \begin{equation}
 \partial_{r_1}\Big(r_1^3\,r_2^a \,(\chi\,\partial_{r_1}\varphi-\varphi\,\partial_{r_1}\chi)\Big)+\partial_{r_2}\Big(r_1^3\,r_2^{a}\,(\chi\,\partial_{r_2} \varphi-\varphi\,\partial_{r_2}\chi) \Big)=[4\,l_1\,(l_1+1)-4\,l_2\,(l_2+1)]\,r_1\,r_2^a\,\varphi\,\chi\, .
 \end{equation}
 So defining the ``current"
 
 \begin{equation}
 j_i=r_1^3\,r_2^a\,(\chi\,\partial_{r_i}\varphi-\varphi\,\partial_{r_i}\chi)\, ,
 \end{equation}
 we can write 
 
 \begin{equation}
 \partial_{r_i}\,j_{r_i}=[4\,l_1\,(l_1+1)-4\,l_2\,(l_2+1)]\,r_1\,r_2^a\,\varphi\,\chi\, .
 \end{equation}
 Hence we find a good candidate for internal product for our wavefunctions, namely
 
 \begin{equation}
 \langle \chi | \varphi\rangle\sim \int r_1\,r_2^a\,\chi\,\varphi\, ,
 \end{equation}
provided we impose boundary conditions such that

\begin{equation}
\label{bc}
\int \partial_{r_i}j_{r_i}=\int_{\partial}\,j_{r_i}\,dr_i=0\, .
\end{equation}
We will be interested on normalizable wavefunctions for which the above inner product is well-defined. This can be done by demanding the current to vanish on the $\alpha=0$ singularity. One can easily check that this is satisfied provided that $a=2\,l$, so that finally, and upon fixing the correct normalization, we have

\begin{equation}
\begin{array}{l c l}
h^{<}_{I}=\frac{\mathcal{C}}{\frac{10}{3}+4\,l}\,\frac{1}{\rho_0^{\frac{10}{3}}}\,\Big(\frac{\rho}{\rho_0}\Big)^{2\,l}\,\cos^{2\,l}\alpha & \leftrightarrow & \rho< \rho_0\, ,  \\
 &  &  \\
h^{>}_{I}=\frac{\mathcal{C}}{\frac{10}{3}+4\,l}\,\frac{1}{\rho^{\frac{10}{3}\,}}\,\Big(\frac{\rho_0}{\rho}\Big)^{2\,l}\,\cos^{2\,l}\alpha & \leftrightarrow & \rho> \rho_0\, . 
\end{array}
\end{equation}
The leading term in the asymptotic region is $\rho^{-\frac{10}{3}}$. Recalling that the $AdS_6$ radial coordinate is $\varrho=\rho^{\frac{2}{3}}$ it is easy to see that, together with the contribution from the overall $\Omega^2$ warping, we recover, in the asymptotic region, the warped $AdS_6\times \hat{S}^4/\mathbb{Z}_2$ geometry. This strongly suggests that this geometry again corresponds to a flow in the original gauge theory triggered by a VEV.

We could again check our results against a computation performed upon changing from the beginning into polar coordinates. Just as in the $AdS_5$ case one easily obtains the same spectrum upon imposing the appropriate quantization conditions on the quantum numbers.

\subsection{Flows on the resolved space}

We now change the $ds_4^2$ metric for that of the Eguchi-Hanson space with resolution parameter $c$. As stressed above, $c\ne 0$ is only possible in the NVS case, as the VS projection kills this mode. The equation for $h$ is now

\begin{equation}
\frac{1}{r^3}\, \partial_{r}\Big(r^3\,f \,\partial_r h \Big)+\frac{1}{z^{\frac{1}{3}}}\,\partial_{z}\Big(z^{\frac{1}{3}}\,\partial_z h \Big)+ \frac{4}{r^2}\,\Delta\,h+\frac{4}{r^2\,f}\,\partial_{\psi}^2\,h=\mathcal{C}\,\delta\, .
\end{equation}
Since we will place our stack of D4 branes on the blow-up $S^2$ where the $\psi$ circle shrinks, we will set $R=0$. Then, expanding $h=\sum_I\,h_I\,\tilde{Y}_I^{\star}(\xi_0)\,\tilde{Y}_I(\xi)$, we find an equation for $h_I$ in the $\{r,\,z\}$ plane. As in the $AdS_5$ case, this equation is fairly complicated, so we will content ourselves with the analysis of the asymptotic properties. Switching to the $\{\rho,\,\alpha\}$ polar coordinates we write

\begin{equation}
h_I=\Big(\frac{c}{\rho}\Big)^a\,f(\alpha)\, .
\end{equation}
We then find the equation

\begin{equation}
(-10 a+3 a^2-6 m^2+a (-10+3 a) \cos(2 \alpha)) f \sin\alpha+\cos\alpha (2 (-4+5 \cos(2 \alpha)) f'+3 \sin(2 \,\alpha)f'')=0\, ,
\end{equation}
where $m^2=4\,l(l+1)$. The solution to this equation is

\begin{equation}
f=\cos^{2l}\alpha\,\,_2F^1[\frac{5}{3}-\frac{a}{2}+l,\,\frac{a}{2}+l,\,2+2\,l,\,\cos^2\alpha]\, .
\end{equation} 
Demanding the current to vanish at $\alpha=0$ sets

\begin{equation}
a=\frac{10}{3}+2\,l\, .
\end{equation}
Hence

\begin{equation}
h_I=\frac{1}{\rho^{\frac{10}{3}}}\,\Big(\frac{c}{\rho}\Big)^{2\,l}\,\cos^{2\,l}\alpha\, .
\end{equation}
Therefore, when written in the $\varrho$ radial coordinate and taking into account the overall warp factor $\Omega$, this geometry is again asymptotically the same $AdS_6\times \hat{S}^4/\mathbb{Z}_2$ cone, thus showing that it must correspond to a VEV deformation of the original --recall, NVS-- CFT.

\subsection{Gauge theory}

Let us now turn to the gauge theory operators. Borrowing the discussion in section \ref{GT4d}, it is clear that the $\tilde{Y}_I$ wavefunctions must correspond to the states in eq.(\ref{states}). 

On the other hand, the two gauge theories relevant to our discussion are, respectively, a $USp(2\,N)\times USp(2\,N)$ gauge theory with bifundamental matter in the VS case and a $SU(2\,N)$ theory with an antisymmetric hypermultiplet in the NVS case. In both cases we can add up to 8 fundamental hypermultiplets. Nevertheless our wavefunctions do not involve any flavor quantum number, signaling that the geometry is blind to flavor degrees of freedom. Recall that in \cite{Bergman:2012qh} the geometry was not able to capture operators on the Higgs branch involving fundamental matter. Hence it comes as no surprise that in this case the same happens. Because of this we will set $N_f=0$.

For future reference, let us spell the symmetries in each case and the representations of each field. In the VS case the global non-R symmetry is $SU(2)_M\times U(1)_{I_1}\times U(1)_{I_2}$, being $SU(2)_M$ a global mesonic symmetry acting on the hypermultiplet. Besides, $U(1)_{I_{1,\,2}}$ are the topological symmetries associated to each gauge group. On the other hand, in the VS case, the global non-R symmetry is $SU(2)_M\times U(1)_B\times U(1)_I$, where again $SU(2)_M$ is a global mesonic symmetry acting on the antisymmetric hypermultiplet. Besides $U(1)_B$ is a baryonic symmetry under which one complex doublet in the antisymmetric hyper  --call it $A_1$-- has charge 1 and the other --call it $A_2$-- has charge -1. Finally $U(1)_I$ is the topological symmetry associated to the gauge group. In addition, in both cases there is a global $SU(2)_R$ symmetry.

It is important to recall that the correct $AdS$ coordinate is $\varrho$. Hence the modes both in the singular and resolved cases scale like $\varrho^{3\,l}$, and thus correspond to $\Delta=3\,l$ operators. Note that indicates no large anomalous dimension, even though, just as in the 4d case, some of the operators taking VEV will be non-chiral (in the 4d sense). This is again due to the combination of $SU(2)_M$ and $SU(2)_R$, which allows to place any operator in the same multiplet as a chiral operator.

\subsubsection{The singular case}

As described above, the geometries corresponding to placing the stack away from the singularity correspond to flows triggered by VEV's of mesonic operators. These flows exist in both VS and NVS theories, which, consequently, have an identical spectrum of mesonic operators \cite{Bergman:2012kr,Bergman:2012qh,Bergman:2013koa}. In fact, in both cases the mesonic moduli space is classified under the $SU(2)_M\times SU(2)_R$ global symmetry, common to both VS and NVS. Using this, since the $\tilde{Y}_I$ correspond to the states in eq.(\ref{states}) whose quantum numbers under all relevant symmetries are known, it is easy to translate among fluctuations and their corresponding operators --basically those listed in \cite{Bergman:2012qh,Bergman:2013koa}-- finding a perfect matching. Exactly as in the 4d case, the $SU(2)_R\times SU(2)_M$ allows to relate non-holomorphic operators to holomorphic ones, hence ensuring that the dimensions should equal the classical ones. Note in particular that $SU(2)_M$ spin $l$ operators involve $2\,l$ scalars --which in 5d are dimension $\frac{3}{2}$, and hence they have the expected $\Delta=3\,l$.

Note that the operators so constructed belong to the Higgs branch. As opposed to the case of 4d gauge theories, in 5d the $R$-symmetry does not contain a $U(1)_r$. In particular this stands for the fact that the scalar $\phi$ in the 5d vector multiplet is a real field. Hence the $n$ quantum number analogous to that in sect. \ref{4dsingular} is absent. Nevertheless one might wonder about operators including $q$ powers of  $\phi$, which in this case seems to be absent. Indeed, the boundary conditions setting the current at $\alpha=0$ to zero don't allow for any other quantum number in any similar way to the $q$ in sect.\ref{4dresolved}. In support of this, analogously to sect.\ref{4dsingularalternative}, one can repeat the computation in polar coordinates from the beginning and make use of angular eigenfunctions on the $S^4$ to impose the correct quantization conditions. However, on the $S^4$ the eigenfunctions are classified into $SO(5)$ reps. The two cartans of $SO(5)$ must correspond to $l_3$ and $R$, so there is no room for another quantum number, in agreement with the discussion above.

\subsubsection{The resolved case}

We now turn to the resolved case. As emphasized above, the resolution mode is only allowed in the NVS case. Hence, the fluctuations we have obtained correspond to the states in eq.(\ref{states}) upon setting $R=0$. 

The lowest dimensional operators taking a VEV is a triplet of scalars involving $\mathcal{U}\sim A_1\,A_1^{\dagger}-A_2\,A_2^{\dagger}$ at $m=0$. These correspond to the scalars in the $U(1)_B$ conserved current multiplet as expected for a flow triggered by a baryonic VEV. 

With no loss of generality we can assume a VEV for the $A_1$ field proportional to $c$. Then, the baryon-like operator $\mathcal{B}=A_1^N$ of dimension $\Delta=\frac{3}{2}\,N$ would acquire a VEV. On the other hand, we can consider an euclidean $D2$ brane wrapping $X=\{r,\,z,\,\psi\}$, which stands for the baryonic condensate \cite{Bergman:2012kr}. The DBI action for the brane is

\begin{equation}
S_{DBI}=i\,\frac{T_2}{2}\,(2\,\pi)\,\Big(\frac{2}{3\,m}\Big)^{-\frac{1}{3}}\,\int_X \,r\,z^{\frac{1}{3}}\,h\, .
\end{equation}

Asymptotically the integrals for the $h_I$ modes are 

\begin{equation}
\int d\rho\,\rho^{-1-2\,l}\, .
\end{equation}
so again the $l>0$ yield finite integrals, while the $l=0$ term leads to a logarithmically divergent term which has to be regulated by a cut-off. It is easy to see that, upon using the correct $AdS$ radial coordinate $\varrho$, the leading divergence goes like $\frac{3}{2}\,N\,\log\varrho_c$, being $\varrho_c$ the UV cut-off. Hence this corresponds to the dimension of the expected baryon operator VEV, namely $\Delta=\frac{3}{2}\,N$. Furthermore, we can consider a $D4$ brane wrapping the blown-up $S^2$ and describing a cosmic string in the field theory directions --which is a $1+2$ dimensional defect in 5d--. Its DBI action is 

\begin{equation}
S_{DBI}=-T_4\,\int \,e^{-\Phi}\,\Omega^5\,h^{-\frac{1}{4}}\,\frac{c^2}{4}\,\sin\theta=-T_4\,c^2\,\pi\,\int dx^0\,dx^1\,dx^2\, .
\end{equation}
So we find a finite-tension object ``electric-magnetic" dual to the baryon condensate. The $\delta F_6$ sourced by the brane is of the form

\begin{equation}
\delta F_6=da_3\wedge W\, ;
\end{equation}
being $W$ a closed 2-form in the internal space which asymptotes to the volume form of the blown-up $S^2$. The dual $\delta F_4$ is of the form 

\begin{equation}
\delta F_4=\star_5da_3\wedge \Big(\Omega^{-2}\,h\,\star_{internal}\,W\Big)\, .
\end{equation}
Writing $\star_5da_3=dp$, a local integration  is

\begin{equation}
\delta C_3=p\wedge \Big(\Omega^{-2}\,h\,\star_{internal}\,W\Big)\, .
\end{equation}
It is clear that $\Big(\Omega^{-2}\,h\,\star_{internal}\,W\Big)$ threads the cycle wrapped by the baryonic condensate, so that its WZ action is proportional to $i\,p$. Therefore we see that indeed the baryon condensate captures the baryonic VEV including the Goldstone boson of the broken symmetry. The D4 brane is nothing but the cosmic string around which the Goldstone boson of the --spontaneously broken-- baryon symmetry winds.

\section{Conclusions}\label{conclusions}

In this paper we have studied geometries dual to flows triggered either by mesonic and baryonic VEV's in gauge theories with 8 supercharges in 4 and 5 dimensions. As opposed to the $\mathcal{N}=1$ flows well studied in the literature such as \textit{e.g.} \cite{Klebanov:2007us}, in this case we need to solve an involved PDE, whose general solution in the resolved cases we have not been able to find. 

At the bottom of the geometries, close to the source branes, we expect an $AdS$ throat to emerge, corresponding to the IR fixed point. Unfortunately, due to the lack of explicit solution to the PDE in the resolved cases, we have not been able to explicitly show it. 

Quite remarkably, even though some of the operators taking VEV's in our flows are non-chiral, the dimensions are those of the free field theory. As explained above, this is due to the $SU(2)_M\times SU(2)_R$ symmetry, whose combination allows to regard any non-chiral operator as in the same multiplet of a chiral operator. It is interesting to note that this seems to be a property of the $\mathbb{Z}_2$ orbifold theories, as for a generic $\mathbb{Z}_p$ orbifold the mesonic symmetry is just $U(1)_M$. We leave a detailed study for the future.

In the 5d case the singularity at $\alpha=0$ plays an important role in providing the correct quantization conditions. Indeed, without such conditions one seems to obtain regular modes elsewhere for an arbitrary $a$ --basically the scaling dimension of the dual operator, since we consider $h_I\sim \rho^{-a}\,f(\alpha)$--. However, demanding the vanishing of the current at $\alpha=0$ yields to the correct $AdS_6$ asymptotics --which demand an overall $\rho^{-\frac{10}{3}}$-- and gives the correct --and discrete-- dimension to the operators. As raised in the text, one slightly puzzling feature is that the 5d operators taking a VEV are purely Higgs branch operators with no vector multiplet scalar insertions. Note that, despite its singularity at $\alpha=0$, the SUGRA background captures well the CFT properties, as it also happens in the case of \cite{Bergman:2012qh}. 

The resolved geometries correspond to flows triggered by VEV's of baryonic operators. Indeed we find a fully consistent picture, with the baryonic VEV being identified with the appropriate baryonic condensate brane. Its DBI action gives the modulus of the VEV with the expected dimensional scaling. While we have not been able to compute the finite, wavefunction, part of the action since we don't have the exact form for the warp factor, we expect a similar result as in \cite{Klebanov:2007us}. In fact, in the $AdS_5$ case, where the internal space is the base of a $CY_3$ cone, we can simply borrow the general result in \cite{Benishti:2010jn} appropriately adapted to the $CY_3$ case. On the other hand, the DBI action for the baryon condensate brane is nicely proportional to the Goldstone boson, sourced by a finite-tension cosmic string brane. While we have not been able to check the existence and properties of the required $W$-form since, for a start, the exact $h_I$ are not known, we believe that it will exist along the lines of \cite{Klebanov:2007cx}.

\section*{Acknowledgements} 

We are grateful to D.Martelli for useful conversations. D.R-G thanks O.Bergman and G.Zafrir for previous collaborations and conversations. The authors are partially supported by the Spanish Ministry of Science and Education grant FPA2012-35043-C02-02, by the Government of Asturias grant SV-PA-12-ECOEMP-30 and by the Ram\'on y Cajal fellowship RyC-2011-07593. They also acknowledge partial support from the University of Oviedo under the grant no. UNOV-12-EMERG-18.

\begin{appendix}

\section{Some explicit details about the $CY_3$ structure of $\mathbb{C}^2/\mathbb{Z}_2\times \mathbb{C}$}\label{geometry}

Being a direct product, the metric on the singular $\mathbb{C}^2/\mathbb{Z}_2\times \mathbb{C}$ can be easily written as

\begin{equation}
\label{metric}
ds_6^2=dX_I\,g_6^{IJ}\,dX_J=dr_1^2+\frac{r_1^2}{4}\,(d\psi+\cos\theta\,d\phi)^2+\frac{r_1^2}{4}\,\Big(d\theta^2+\sin^2\theta\,d\phi^2\Big)+dr_2^2+r_2^2\,d\chi^2\, ,
\end{equation}
where $\psi\,\in\,[0,\,2\,\pi]$. Note that, upon introducing $r_1=\rho\,\cos\alpha$, $r_2=\rho\,\sin\alpha$ the metric becomes just $d\rho^2+\rho^2\,ds_{S^5/\mathbb{Z}_2}^2$

We can write the metric on the resolved $\mathbb{C}^2/\mathbb{Z}_2\times \mathbb{C}$ simply by plugging the Eguchi-Hanson metric in the $\mathbb{C}^2/\mathbb{Z}_2$ piece as

\begin{equation}
\label{metricr}
ds_6^2=\frac{dr_1^2}{f(r_1)}+\frac{r_1^2}{4}\,f(r_1)\,(d\psi+\cos\theta\,d\phi)^2+\frac{r_1^2}{4}\,\Big(d\theta^2+\sin^2\theta\,d\phi^2\Big)+dr_2^2+r_2^2\,d\chi^2\, ,
\end{equation}
being

\begin{equation}
f(r_1)=1-\frac{c^4}{r_1^4}\, .
\end{equation}
Introducing complex coordinates $\{z_1,\,z_2,\,z_3\}$ defined as

\begin{equation}
z_1=(r_1^4-c^4)^{\frac{1}{4}}\,e^{i\frac{\psi+\phi}{2}}\,\cos\frac{\theta}{2}\, ,\qquad z_2=(r_1^4-c^4)^{\frac{1}{4}}\,e^{i\frac{\psi-\phi}{2}}\,\sin\frac{\theta}{2}\, ,\qquad z_3=r_2\,e^{i\chi}\, ,
\end{equation}
it is easy to check that the K\"{a}hler potential reads

\begin{equation}
F=c^2\,\sqrt{1+\frac{z_1\,\bar{z}_1+z_2\,\bar{z}_2}{c^4}}-c^2\,\log\Big(1+\sqrt{1+\frac{z_1\,\bar{z}_1+z_2\,\bar{z}_2}{c^4}}\Big)+c^2\,\log\Big(1+\frac{z_1\,\bar{z}_1}{z_2\,\bar{z}_2} \Big)+z_3\,\bar{z}_3\, .
\end{equation}
Furthermore, introducing the natural complex 1-forms 

\begin{equation}
e_1=\frac{dr_1}{\sqrt{1-\frac{c^4}{r_1^4}}}+i\,\frac{r_1}{2}\,\sqrt{1-\frac{c^4}{r_1^4}}\,g_5\, ,\qquad e_2=\frac{r_1}{2}\,(d\theta-i\,\sin\theta\,d\phi)\, ,\qquad e_3=dr_2+i\,r_2\,d\chi\, ,
\end{equation}
where $g_5=d\psi+\cos\theta\,d\phi$, one can easily verify that the K\"{a}hler form arising from the K\"{ a}hler potential is just $\omega=\sum_{i=1}^3e_i\wedge\bar{e}_i$. The holomorphic 3-form is

\begin{equation}
\Omega=e^{i\,(\psi+\chi)}\,e_1\wedge e_2\wedge e_3\, .
\end{equation}
One can easily verify that

\begin{equation}
d\Omega=0\qquad d\omega=0\qquad \Omega\wedge\omega=0\, .
\end{equation}

This explicitly shows the $CY_3$ structure of the resolved $\mathbb{C}^2/\mathbb{Z}_2\times \mathbb{C}$.

\end{appendix}

\end{document}